\documentclass[twocolumn,letter]{jpsj2} 
\title{An Improved Initialization Procedure for the Density-Matrix Renormalization Group}

\author{Masaki \textsc{Tezuka}
\thanks{tezuka@cms.phys.s.u-tokyo.ac.jp}}

\inst{
Department of Physics, University of Tokyo\\
Hongo, Bunkyo-ku, Tokyo 113-0033}

\abst{We propose an initialization procedure
for the density-matrix renormalization group (DMRG): {\it the recursive sweep method}.
In a conventional DMRG calculation,
the infinite-algorithm, where two new sites are added to the system at each step,
has been used to reach the target system size.
We then need to obtain the ground state for a different system size
for every site addition, so 1) it is difficult to supply
a good initial vector for the numerical diagonalization for the ground state,
and 2) when the system reduced to a 1D system consists of an
array of nonequivalent sites as in ladders or Hubbard-Holstein model,
special care has to be taken.
Our procedure, which we call
the {\it recursive sweep method}, provides a solution to these problems and
in fact provides a faster algorithm for the Hubbard model as well as more
complicated ones such as the Hubbard-Holstein model.}

\kword{DMRG, infinite algorithm DMRG, finite algorithm DMRG, exact diagonalization,
ladder, Hubbard model, electron-phonon interaction, phonon,
Hubbard-Holstein model}

\begin{document}
\maketitle
\paragraph{Introduction}
After fifteen years since
the density-matrix renormalization group (DMRG)
\cite{White1992PhRvL..69.2863W, White1993PhRvB..4810345W}
was introduced as a new numerical method
to obtain ground- and low-energy excited states for one-dimensional
many-body systems,
DMRG continues to enjoy its status as one of the most powerful numerical tools
in condensed matter physics.
It has been successfully applied to study various problems \cite{1998DMRGbook},
which include
correlated electron systems such as the Hubbard model
\cite{PhysRevLett.73.882, PhysRevB.53.R10445, PhysRevB.57.10324}
and the $t$-$J$ model \cite{1998PhRvL..80.1272W},
quantum systems at finite temperatures \cite{0953-8984-8-40-003,JPhysSocJpn.66.2221,PhysRevB.56.5061},
quantum chemistry \cite{2002JChPh.117.7472W},
quantum Hall systems \cite{PhysRevLett.86.5755},
dynamical quantities \cite{PhysRevB.52.R9827, PhysRevB.60.335, PhysRevB.66.045114}.
The extreme accuracy obtained by DMRG in many cases has attracted
interests for its mathematical reasons, and such viewpoint is also
pursued vigorously
(see, {\it e.g.}, \cite{2004PhRvL..93v7205V,2004PhRvB..70t5118L}).
One of the recent breakthroughs in the field is the development of
several new methods to apply DMRG to real-time simulations
of quantum systems \cite{2002PhRvL..88y6403C,
2004PhRvL..93d0502V,2004PhRvL..93g6401W}.
Several review articles
\cite{2005RvMP...77..259S, 2006cond.mat..3842D,2006cond.mat..9039H}
describe the history of DMRG and its wide range of application.

\begin{figure}
\begin{center}
\includegraphics[width=7cm]{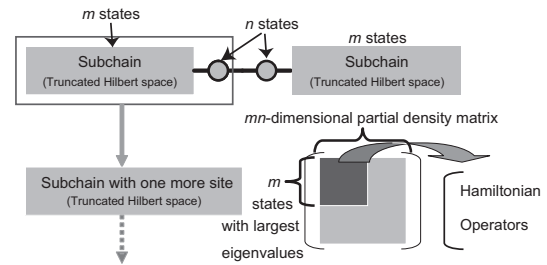}
\caption[DMRG calculation.]{DMRG calculation.
The one-dimensional lattice model is treated by
iteratively adding sites to two subchains to construct
superblocks from two subchains and two sites between them.
In enlarging a subchain to contain one more site,
the partial density matrix for the new block is calculated for
the target state of the superblock.
The eigenstates with $m$ largest eigenvalues are retained
as the new basis for the enlarged subchain.
The truncation error is assessed by calculating the sum $\eta$ of
the eigenvalues for the discarded eigenstates.
In the new basis, the matrix representations of the Hamiltonian and
operators for the new block are calculated and stored for later use.
}
\label{fig:DMRGstep}
\end{center}
\end{figure}

In DMRG, we consider a system consisting of sites aligned in
a one-dimensional open-boundary chain.
We systematically transform the Hilbert space of the system
to one whose dimension is drastically reduced from the original one,
while keeping the low-energy physics as unchanged as possible.
In order to do this, we separate the system into parts,
which we call {\it blocks} or {\it subchains}, and reduce the number of basis functions
to describe the blocks, which we call the {\it basis size} for the blocks.
What we actually do is to choose a basis that is fit to express
one (or a few) quantum-mechanical state(s) of the original system ({\it target state(s)}),
and expect that the low-energy excitations are recaptured in the chosen basis.
The basis size to express physical states
is reduced with the use of a partial density matrix
calculated for some parts of a {\it superblock},
which is the whole or a partial chain and
is usually constructed from two blocks and two sites.
The part to calculate the partial density matrix is
separated at some {\it cut} point from the rest of the system,
which is called the {\it environment block}.
While the number of states to be considered in an exact diagonalization
increases exponentially with the number of sites,
the contribution of high-energy configurations to the low-energy physics
is usually exponentially small.
With the use of the DMRG algorithm, most important states can be
systematically found. In many cases,
the dimension of the Hilbert space can be kept much smaller than
in the exact diagonalization, and
almost exact results can be obtained with moderate computational efforts.

In DMRG we need to start with a warm-up process, which is the process to iteratively
add new sites to one or more subchains, until
the Hilbert space of the system with the desired number of sites
can be represented with the reduced basis.
If the system consists of homogeneous sites,
we can just add two new sites between two subchains of the same length $l$
to make a superblock of $2l+2$ sites
in order to obtain a subchain that has $l+1$ sites,
and repeat this until the superblock acquires the desired number of sites.
This process is called the {\it infinite algorithm DMRG};
we can repeat the addition of a site to the subchain indefinitely.
Once this warm-up procedure has finished,
we can enhance the quality of the basis
by systematically moving the cut location back and forth
within the chain
(the {\it finite algorithm DMRG}).

In this letter, we focus on the warm-up process.
When we increase the number of sites by two in the infinite-algorithm,
it is in general difficult to supply a good initial vector for the
numerical diagonalization for the target state,
so usually a random vector is plugged.
Our new procedure reduces the number of numerical diagonalization
from a random vector, with the use of the finite algorithm DMRG.
\paragraph{Method}
We illustrate the new procedure for the case of $N_c$ original sites per full site.
Our idea here is to modify the infinite algorithm in the conventional DMRG,
which adds two sites at the center of the superblock at each step,
so that two full sites are added at the center of the superblock per cycle.
As we elaborate in the below, for each of the left and right subchains,
we retain $N_c$ subchains
that have respectively $0,1,\ldots,N_c-1$ original sites besides $n$ full sites,
instead of just one subchain each. The longest subchain among the left subchains
and the longest among the right ones are used to construct
a new superblock, and with the use of the finite algorithm,
those subchains are enlarged by $1,2,\ldots,N_c$ original sites.
Then we can repeat the same process for a superblock that is two full sites longer.

Now we explain how this can be done.
We align the original sites in a one-dimensional chain
so that the same kind of sites appear periodically along the chain, as 
\[
\cdots \mbox{\textsf{) --- (b}}_0\mbox{\textsf{---}}
\cdots \mbox{\textsf{---b}}_{N_c-1}\mbox{\textsf{) --- (b}}_0\mbox{\textsf{---}}
\cdots \mbox{\textsf{---b}}_{N_c-1}\mbox{\textsf{) --- (}}\cdots,
\]
where the original sites are denoted as \textsf{b}$_0$,
\ldots,\textsf{b}$_{N_c-1}$,
and each full site is enclosed in \textsf{(}$\cdots$\textsf{)}.

\begin{figure}[h]
\begin{center}
\includegraphics[width=7cm]{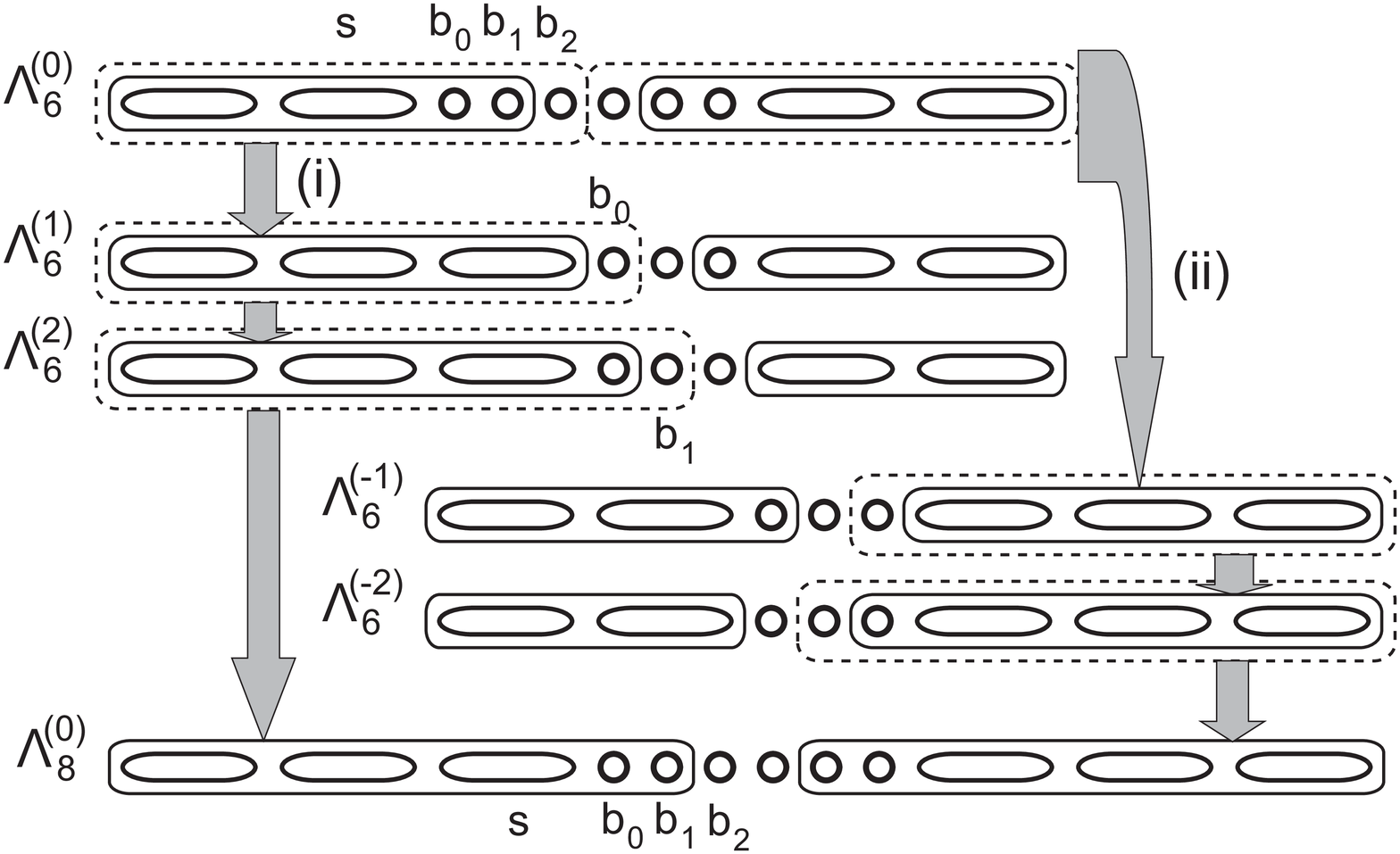}
\caption[]
{The recursive sweep method illustrated for the case of $n=2$ and
$N_c=3$.
A full site, \textsf{s}, is represented as an oval,
and an original site, $\textsf{b}_i$, as a circle.
An arrow shows how an original site is added to
a subchain, which is represented as a round cornered rectangle.
The first steps to enlarge the left and right subchains are marked as
(i) and (ii), as in the text.
For the definition of the superblocks $\Lambda_6^{(0,\pm1,\pm2)}$
and $\Lambda_8^{(0)}$, see the text.
}
\label{fig:recursivesweep}
\end{center}
\end{figure}

We denote the full site as \textsf{s},
and a sequence of $n$ full sites as \textsf{s}$^n$.
Suppose we have the left and right subchains $L_{n,i}$ and $R_{n,i}$ ($i=0,1,\ldots,N_c-1$)
respectively,\\
$L_{n,0} = \textsf{s}^n$,
$R_{n,0} = \textsf{s}^n$,\\
$L_{n,1} = \textsf{s}^n$\textsf{---b}$_0$,
$R_{n,1} =$ \textsf{b}$_{N_c-1}$\textsf{---}$\textsf{s}^n$,\\
$L_{n,2} = \textsf{s}^n$\textsf{---b}$_0$\textsf{---b}$_1$,
$R_{n,2} =$ \textsf{b}$_{N_c-2}$\textsf{---b}$_{N_c-1}$\textsf{---}$\textsf{s}^n$,\\
\ldots,\\
$L_{n,N_c-1} = \textsf{s}^n$\textsf{---b}$_0$\textsf{---}$\cdots$\textsf{---b}$_{N_c-1}$,
and\\
$R_{n,N_c-1} =$ \textsf{b}$_1$\textsf{---}$\cdots$\textsf{---b}$_{N_c-2}$\textsf{---b}$_{N_c-1}$\textsf{---}$\textsf{s}^n$,\\
where $L_{n,i}$ ($R_{n,i}$) has
 (a) $n$ full sites from the left(right) edge of the original chain
and (b) $i(=0,1,\ldots,N_c-1)$ original sites from
the $(n+1)$-th site $s_n$ ($s_{L-n-1}$).
We enlarge $L_{n,N_c-1}$ and $R_{n,N_c-1}$ by $N_c$ original sites
to obtain $2N_c$ subchains $L_{n+1,i}$ and $R_{n+1,i}$ ($i=0,1,\ldots,N_c-1$)
as follows.

First, we construct a superblock $\Lambda_{2n+2}^{(0)}$ that has $2n+2$ full sites:
\[
\Lambda_{2n+2}^{(0)} = L_{n,N_c-1}\mbox{\textsf{---b}}_{N_c-1}(p_\alpha)
\mbox{\textsf{---b}}_0(p_\beta)\mbox{\textsf{---}}R_{n,N_c-1}
\]
is constructed from $L_{n,N_c-1}$, two sites $p_\alpha$
and $p_\beta$, and $R_{n,N_c-1}$, aligned in this order from the left.
We calculate the target state(s) for $\Lambda_{2n+2}^{(0)}$.
Then we can obtain
(i) $L_{n+1,0}$ by calculating the partial density matrix for $L_{n,N_c-1}$ and $p_\alpha$ combined, and
in the same way, (ii) $R_{n+1,0}$ from $p_\beta$ and $R_{n,N_c-1}$ combined.
We write this as
\[
\mbox{(i) \sf[}L_{n,N_c-1}\mbox{\textsf{---b}}_{N_c-1}(p_\alpha)\mbox{\textsf{]}}
\rightarrow L_{n+1,0},
\]
and
\[
\mbox{(ii)} R_{n+1,0}\leftarrow \mbox{\textsf{[b}}_0(p_\beta)\mbox{\textsf{---}}R_{n,N_c-1}\mbox{\textsf{]}}.
\]

Next, as in the finite algorithm,
we move the cut location to the right by one original site.
This is possible because we have $R_{n,N_c-2}$.
The new superblock, $\Lambda_{2n+2}^{(1)}$ also has $2n+2$ full sites:
\[
\Lambda_{2n+2}^{(1)} = L_{n+1,0}\mbox{\textsf{---b}}_0(p_\beta)
\mbox{\textsf{---b}}_1(p_\gamma)\mbox{\textsf{---}}R_{n,N_c-2}
\]
is constructed from $L_{n+1,0}$, two original sites $p_\beta$ and $p_\gamma$,
and $R_{n,N_c-2}$.
Then we obtain $L_{n+1,1}$ by calculating the partial density matrix for $L_{n+1,0}$ and
$p_\beta$ combined:
\[
\mbox{\textsf{[}}L_{n+1,0}\mbox{\textsf{---b}}_0(p_\beta)\mbox{\textsf{]}}
\rightarrow L_{n+1,1}.
\]
This can be repeated $N_c-2$ more times,
where for each $i$ ($i=2,\ldots,N_c-1$), we obtain $L_{n+1,i}$
from the superblock $\Lambda_{2n+2}^{(i)}$
that is constructed from $L_{n+1,i-1}$, two original sites, and $R_{n,N_c-1-i}$:
\[
\Lambda_{2n+2}^{(i)} = L_{n+1,i-1}\mbox{\textsf{---b}}_{i-1}
\mbox{\textsf{---b}}_{i}\mbox{\textsf{---}}R_{n,N_c-1-i},
\]
\[
\mbox{\textsf{[}}L_{n+1,i-1}\mbox{\textsf{---b}}_{i-1}\mbox{\textsf{]}}
\rightarrow L_{n+1,i}.
\]

Also, by moving the cut location to the left $N_c-1$ times,
we make superblocks $\Lambda_{2n+2}^{(-i)}$ ($i=1,\ldots,N_c-1$) that consists 
of $L_{n+1,N_c-1-i}$, two original sites and $R_{n,i-1}$,
to obtain $R_{n+1,i}$ ($i=1,\ldots,N_c-1$):
\[
\Lambda_{2n+2}^{(-i)} = L_{n,N_c-1-i}\mbox{\textsf{---b}}_{N_c-1-i}
\mbox{\textsf{---b}}_{N_c-i}\mbox{\textsf{---}}R_{n+1,i-1},
\]
\[
R_{n+1,i}\leftarrow \mbox{\textsf{[b}}_{N_c-i}\mbox{\textsf{---}}R_{n+1,i-1}\mbox{\textsf{]}},
\]
for $i=1,\ldots,N_c-1$.

In these steps, the initial vector for each of the exact diagonalizations can be calculated
from the target state obtained in the most recent step
(for $\Lambda_{2n+2}^{(-1)}$, we use the target state for $\Lambda_{2n+2}^{(0)}$,
rather than $\Lambda_{2n+2}^{(+N_c-1)}$).
The calculations for $\{L_{n+1,i}\}$ and $\{R_{n+1,i}\}$
can be done without data dependence to each other in a parallel computer.

Now we have
$L_{n+1,i}$ ($i=0,\ldots,N_c-1$) and
$R_{n+1,i}$ ($i=0,\ldots,N_c-1$), 
so we have enlarged the subchains by one full site.
We have to obtain the target states from random vectors only once.
This procedure can be repeated for the incremented values of $n$,
until the superblock has all the $L$ full sites.

Note that, because DMRG limits the number of the basis functions to $m$
to describe finite-size blocks,
a calculation with any choice of basis should converge to the result
of the exact diagonalization in the $m\rightarrow\infty$ limit.
The questions here are:
\begin{itemize}
\item how fast the calculation converges, and
\item whether we can calculate with enough accuracy
within a practical computation time.
\end{itemize}

We also note that the recursive sweep method described here is different from
(a) the initialization by Liang and Pang \cite{PhysRevB.49.9214}
for a rectangle lattice where the ratio of horizontal
and vertical hopping parameters is controlled through several finite-algorithm sweeps
that involve the whole system,
(b) the warm-up with a conventional numerical renormalization group
\cite{1975RvMP...47..773W}
by Xiang \cite{PhysRevB.53.R10445},
the {\it two-step DMRG} by Moukouri and Caron \cite{2003PhRvB..67i2405M},
and
(c) the quantum information entropy-based approach by Legeza and S\'olyom
\cite{2003cond.mat..5336L}.

\subparagraph{Choice of targets}
Because environment blocks are shorter than the enlarged block in
the finite-algorithm steps, it is advisable to target several states
having different quantum numbers ({\it e.g.}, numbers of
up-spin electrons and down-spin electrons)
near the target state.
In calculating for the half-filled, $L$-site system ($L$: an even integer),
we target not only the ground state for $(n_\uparrow,n_\downarrow)=(L/2,L/2)$
but also the ground states for
$(n_\uparrow+d_{\uparrow,j}, n_\downarrow+d_{\downarrow,j'})$
($d_j,d_{j'}=0,\pm1$).

\begin{figure}[h]
\begin{center}
\includegraphics[width=7cm]{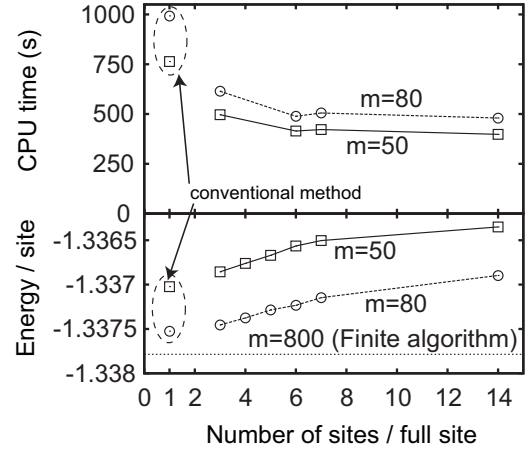}
\caption{
Upper panel: The CPU time for the warm-up for the 84-site
Hubbard model with $U/t=2$, 
for the number of retained eigenstates $m=30,50,80$,
as function of the number of sites in a full site
for the recursive sweep method.
The calculations were conducted on a workstation
with two Xeon 5160 chips and 4GB of RAM.
Lower panel: The ground-state energy per site at the end of
the warm-up for the 84-site Hubbard model with $U/t=2$, 
for the number of retained eigenstates $m=50,80$,
as function of the number of sites in a full site
for the recursive sweep method.
The result for the conventional infinite method is also shown.
The true energy, which can be estimated with the finite-algorithm
DMRG with a large $m=800$, is $-1.337~785~42$
and shown as a dotted line.
Other lines are guide to the eye.
The ground state energy per site for the {\it infinite-length}
Hubbard model with $U=2$, obtained with the exact solution
by Lieb and Wu, is
$-1/2 -\int_{-\infty}^\infty J_0(x)J_1(x)/x(1+e^{|x|}){\rm d} x= -1.344~374~34$.
}
\label{fig:inffinenergy_m}
\end{center}
\end{figure}
\paragraph{Results}
\subparagraph{The Hubbard model}
We first apply the recursive sweep method to the half-filled Hubbard chain
with homogeneous sites,
\begin{equation}
H=-\sum_{i,\sigma}t(c_{i,\sigma}^\dag c_{i+1,\sigma} + \mbox{H.c.})
+ \sum_i U (n_{i,\uparrow}-1)(n_{i,\downarrow}-1),
\label{eqn:Hubbard}
\end{equation}
by considering multiple sites as one composite {\it full site}.
In each iteration of the recursive sweep method,
we add two composite sites to the superblock.
When the length of the superblock equals or exceeds the desired length $L_0$,
we can stop the recursive sweep method and start the finite-algorithm sweeps.

As shown in Figure \ref{fig:inffinenergy_m},
the time required for the warm-up stage is almost halved
in the case of the 84-site Hubbard model with $U/t=2$
by the use of this method.
The longer cycle we take, generally the faster the calculation becomes.
We also plot the ground state energy per site
in Figure \ref{fig:inffinenergy_m}.
The ground state energy becomes slightly worse than in the conventional
infinite-algorithm warm-up with the same number of retained states $m=50$.
When we increase $m$ to $m=80$, however, the energy becomes better
than in the infinite-algorithm with $m=50$ for up to seven sites per cycle,
while the CPU time spent is still significantly lower.
So a small increase in $m$ can compensate the increased error due to
the use of the recursive sweep algorithm with $N_c=2-7$.

\subparagraph{The Hubbard-Holstein model}
When the system is a repetition of three or more types of sites,
as in the case of the pseudo-site method for the Holstein phonons
\cite{Jeckelmann1998PhRvB..57.6376J}
or multi-leg ladder systems,
the choice of the warm-up process in DMRG is not obvious.
Hereafter, we call one original site, which becomes the
period of the repetition of the (pseudo-)sites,
as a {\it full site}.
(a) If we enlarge the two subchains in a mirror-symmetric way,
at most of the steps, we have two incomplete full sites at the center
of the superblock.
Then the environment for the new sites at the center
is much different from the later stages of the calculation,
when the full site is completely within the superblock.
(b) Alternatively,
we can increase the length of the subchain that starts from one edge of
the chain by iteratively using short subchains whose whole basis
can be exactly treated, in a way in which
the sites in the superblock constitute a number of full sites at each step.
In this case the new sites are always added near the
edge of the chain, even when they are around the center of the original system.
So both of the approaches (a)(b) have shortcomings that should
deteriorate the choice of the reduced basis,
which can be overcome with the use of the recursive sweep method as follows.

Here we consider the Hubbard-Holstein model \cite{2005PhRvL..95v6401Tr} on a one-dimensional chain,
\begin{equation}
\begin{split}
H&=-\sum_{i,\sigma}t(c_{i,\sigma}^\dag c_{i+1,\sigma} + \mbox{h.c.})
+ \sum_i U n_{i,\uparrow} n_{i,\downarrow}\\
&+ \sum_{i,\sigma} g n_{i,\sigma}(a_i+a_i^\dag)
+ \sum_i \hbar\omega_0 a_i^\dag a_i.
\label{eqn:HHorig}
\end{split}
\end{equation}
Here,
$c_{i,\sigma}$ annihilates an electron with spin $\sigma(=\uparrow,\downarrow)$ at site $i$,
$n_{i,\sigma}=c_{i,\sigma}^\dag c_{i,\sigma}$ is the electron number,
$g$ is the electron-phonon coupling, and
$a_i$ is the phonon annihilator at site $i$.
For the application of the pseudo-site method to the Hubbard-Holstein model,
the present author, in a collaboration with Arita and Aoki \cite{2005PhyB..359..708T},
has previously come up with another method, the {\it compensation method},
to make improve the choice of basis for the infinite-algorithm warm-up.
Because DMRG is a variational method, the calculated ground state energy
is always higher than the actual value; the lower energy means the better convergence.

\begin{figure}[h]
\begin{center}
\includegraphics[width=8cm]{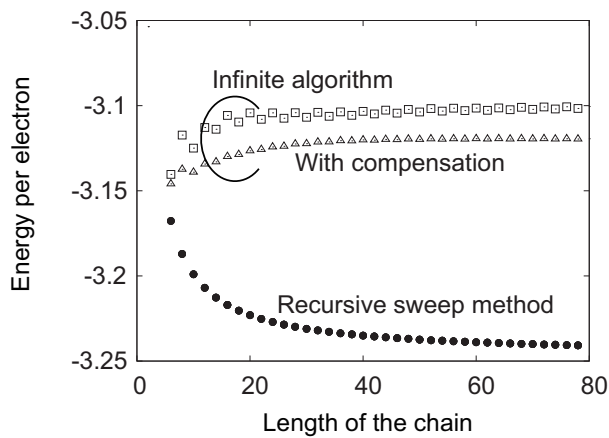}
\caption[]
{The energy per electron in units of $t$
against the length of the chain at intermediate stages of the warm-up
process, calculated
(i) with the infinite algorithm without the compensation method,
(ii) with the infinite algorithm with the compensation method,
and (iii) with the recursive sweep method,
for the Hubbard-Holstein model
with $(U/t,g/t,\omega_0/t)=(2,3,5)$.
$2^{N_b}$=16 is chosen for the cutoff in the phonon number.
The size of the basis is $m=70$ for all the calculations.
The memory required for the calculation
is around 700MB for the recursive sweep method,
when all subchains and operators for calculation of various correlation functions
are retained for later use.}
\label{fig:pseudochange}
\end{center}
\end{figure}

As we show in Figure \ref{fig:pseudochange},
the recursive sweep method warm-up (solid circle in the plot)
results in a much better energy per site
that does not fluctuate and continues to decrease as the number of full sites
in the left subchain is increased.
Similar results have also been observed for other parameter sets.

\paragraph{Summary}
The recursive sweep algorithm allows an initialization of a DMRG calculation
when the system is a repetition of multiple kinds of sites,
without the need of treating incomplete cycles of sites or
the need of adding new sites at locations much off the center of a superblock.
It is an extention of the infinite algorithm DMRG by the finite algorithm sweeps,
over one cycle of added sites to both directions along the chain, 
whose computational cost is much smaller than in the conventional infinite algorithm.
We have demonstrated that the present algorithm
improves the calculation for a system of local phonons coupled to correlated electrons
compared to our previous method.
The recursive sweep algorithm is also applicable to a chain with homogeneous sites.
We have demonstrated that we can both reduce the calculation time
and improve the energy when we apply the algorithm with a small increase
in the number of retained states.

\paragraph{Acknowledgment}
The author thanks Prof. Hideo Aoki and Dr. Ryotaro Arita
for many helpful discussions.
\makeatletter
\let\jnl@style=\rm
\def\ref@jnl#1{{\jnl@style#1}}
\def\prb{\ref@jnl{Phys.~Rev.~B}}        
\def\prl{\ref@jnl{Phys.~Rev.~Lett.}}    
\def\jcp{\ref@jnl{J.~Chem.~Phys.}}      
\makeatother

\end{document}